# Uranus Flagship Science-Driven Tour Design: Community Input Poll

Amy Simon (NASA GSFC), Ian Cohen (Johns Hopkins University Applied Physics Lab), Matthew Hedman (University of Idaho), Mark Hofstadter (Jet Propulsion Laboratory), Kathleen Mandt (NASA GSFC), Francis Nimmo (U.C. Santa Cruz)

**Introduction:** The highest priority Planetary Flagship for the decade 2023-2032 is the Uranus Orbiter and Probe (UOP) [1]. This mission enables broad, cross-disciplinary science of an ice giant system, including the planet, moons, rings, and magnetosphere; the ice giants represent a class of planet that is largely unexplored but that is common (in size) among extrasolar planet systems. At the same time, Uranus is unique, possessing a regular satellite system, rings, and a complex magnetosphere surrounding a planet tilted over for extreme seasonal variation. The uniqueness of the Uranian system led to a broad suite of science objectives that can be met through remote sensing, *in situ* atmospheric probe, and fields and particles experiments, Table 1, as determined by the Giant Planet Systems and Ocean Worlds panels [2].

The UOP Decadal design study used a *notional* payload and measurements to scope out the needs for each science Objective and to find a tour that fit within mass, power, and data volume constraints. Due to the rapid nature of that study, the orbital tour could not be perfectly optimized. Instead, a best effort was made to include orbit insertion, probe release and relay, and multiple satellite flybys [2]. Some Objectives were necessarily better met than others by this tour. More recent work [3,4] has shown that this tour can be modified to meet individual science disciplines in a much more efficient manner, particularly for the satellite flybys. In fact, the mission's tour trajectory can be optimized to meet any of the Decadal Science Objectives, by focusing on a particular system target, for example magnetosphere, satellites, or planet interior [3].

**Table 1**. UOP Decadal Science Objectives and Measurements [from 2]

| Discipline | Science Objective | Measurement |
|---|---|---|
| Atmospheres | **A1. How does atmospheric circulation function, from interior to thermosphere, in an Ice Giant?** | A. Cloud top zonal and 2D winds, waves to ~10 m/s resolution |
| | | B. *In situ* vertical wind profile to 10-20 m/s resolution |
| | | C. Resolved composition, disequilibrium species mapping (to P < 3 bars): $CH_4$, $H_2S$, $H_3^+$, $C_2H_2$, $C_2H_6$, etc., hydrogen ortho/para fraction to mixing ratio ±20% |
| | | D. Depth of atmospheric winds (gravity moments) |
| | **A2. What is the 3D atmospheric structure in the weather layer?** | A. Cloud tomography and aerosols |
| | | B. Vertical temperature profile to ±1K |
| | | C. Global temperature variations in troposphere, stratosphere, thermosphere |
| Atmospheres / Interiors | **A3./I1. When, where, and how did Uranus form, and how did it evolve both thermally and spatially, including migration?** | A. Noble gas (& isotopes of He, Xe) abundances to ± 5% |
| | | B. Elemental (& isotopes of H, C, S, N & O (stretch goal)) abundances, lower bounds on $CH_4$, $H_2S$, $NH_3$, $H_2O$, and the variation with depth |
| | | C. Global distribution of atmospheric composition |
| | | D. Global energy balance (Bond albedo and thermal emission) to 1% |
| Interiors | **I2. What is the bulk composition and its depth dependence?** | A. Gravity field to at least $J_8$, uncertainties on $J_2$-$J_6$ |
| | **I3. Does Uranus have discrete layers or fuzzy core, and can this be tied to its formation and tilt?** | A. Gravity field to at least $J_8$, uncertainties on $J_2$-$J_6$ |
| | | B. Ring oscillations |
| | **I4. What is the true rotation rate of Uranus, does it rotate uniformly, and how deep are the winds?** | A. Internal magnetic field structure |
| | | B. Planet shape |
| Magnetospheres | **M1. What dynamo process produces Uranus's complex magnetic field?** | A. Internal magnetic field structure |
| | **M2. What are the plasma sources & dynamics of Uranus's magnetosphere and how does it interact with the solar wind?** | A. Particles & fields over range of space (distance, longitude, latitude, local time) and time (spin, solar wind variability) |
| | **M3. How does the magnetosphere interact with Uranus's upper atmosphere and satellite surfaces?** | A. Energetic particle fluxes at satellite orbital ranges |
| | | B. Plasma/energetic particle fluxes over Uranus polar regions |
| Rings and Small Satellites | **R1. What processes sculpted the ice giant rings and small moons into their current configuration?** | A. Fine-scale structures in the dense rings at multiple times and longitudes |
| | | B. Measure longitudinal variations in the ring structure (including normal modes and arcs) |
| | | C. Inventory and shape of small moons > 0.5 km in radius within 500,000 km of the planet's center |
| | **R2. What are the compositions, origins and history of the Uranian rings and inner small moons?** | A. Ring (color) imaging at a wide range of phase angles |
| | | B. Ring and small moon spectra (Cordelia to Mab), 1-5 μm |
| Large Satellites and Ocean Worlds | **S1. What are the internal structures and rock-to-ice ratios of the large Uranian moons? Which ones possess substantial internal heat sources or possible oceans?** | A. Magnetic field intensity and direction |
| | | B. Static gravity coefficients |
| | | C. Global shape |
| | | D. Energy distribution of bulk plasma flow, 10eV-10 keV |
| | | E. Plume/activity searches |
| | | F. Satellite orbital positions |
| | **S2. How do the compositions and properties of the Uranian moons and ring system constrain their formation and evolution?** | A. Reflectance spectra from 0.8-5 μm, detect features 1% of continuum from 0.8-2.6 μm and 2% of continuum from 2.6-5.0 μm |
| | | B. Static gravity coefficients |
| | | C. Global shape |
| | **S3. What geological history and processes do the surfaces record and how can they inform outer solar system impactor populations?** | A. Distribution and topography of surface features |
| | | B. Variations in surface composition |
| | | C. Energy distribution of bulk plasma flow, 1eV-1 keV |
| | | D. High-phase plume-search images |
| | **S4. What evidence of exogenic interactions do the surfaces contain?** | A. Energy distribution of bulk plasma flow, 10keV-10 MeV |
| | | B. Variations in surface composition in reflectance spectra |
| | | C. Evidence of radiation processing of surface ices |

The grassroots poll described here was designed to collect broad community inputs on the tour parameters required for *every* Decadal Science Objective; it was a community-led effort with no institutional or NASA sponsorship. Here we present results aggregated by Objective, as well as compared longitudinally, of the highest priority tour needs identified across all Objectives. Both the raw and aggregated data are publicly available through Zenodo for future white papers, tour and trajectory work, as well as mission and instrument design teams, but the results are not binding nor decisional in any way.

**Poll Details:** Our UOP tour poll was conducted from January 3 to March 30, 2025, with notices sent to multiple planetary science mailing lists and posted at the annual Lunar and Planetary Science Conference (LPSC). The poll asked respondents to pick a single Objective from Table 1 and to rank the aspects and characteristics of a potential UOP orbital tour needed to meet that Objective, most with a 1 (lowest priority) to 5 (highest priority) scale. Some questions were specific to satellites or magnetospheres, but all questions were optional if the respondent did not have a strong preference. Participants could choose to submit more than one response, if they wanted to provide input on multiple Objectives. There were 72 responses to the poll, the majority for satellite Objectives, which is unsurprising given the LPSC audience. There were no responses for Objectives I4 or M1. It is important to note that the number of responses is not indicative of science priority, importance, or interest, it simply represents a snapshot in time for those who chose to respond. Figures 1 and 2 show the full set of poll questions, asked for every Objective (questions are also available in the data spreadsheets).

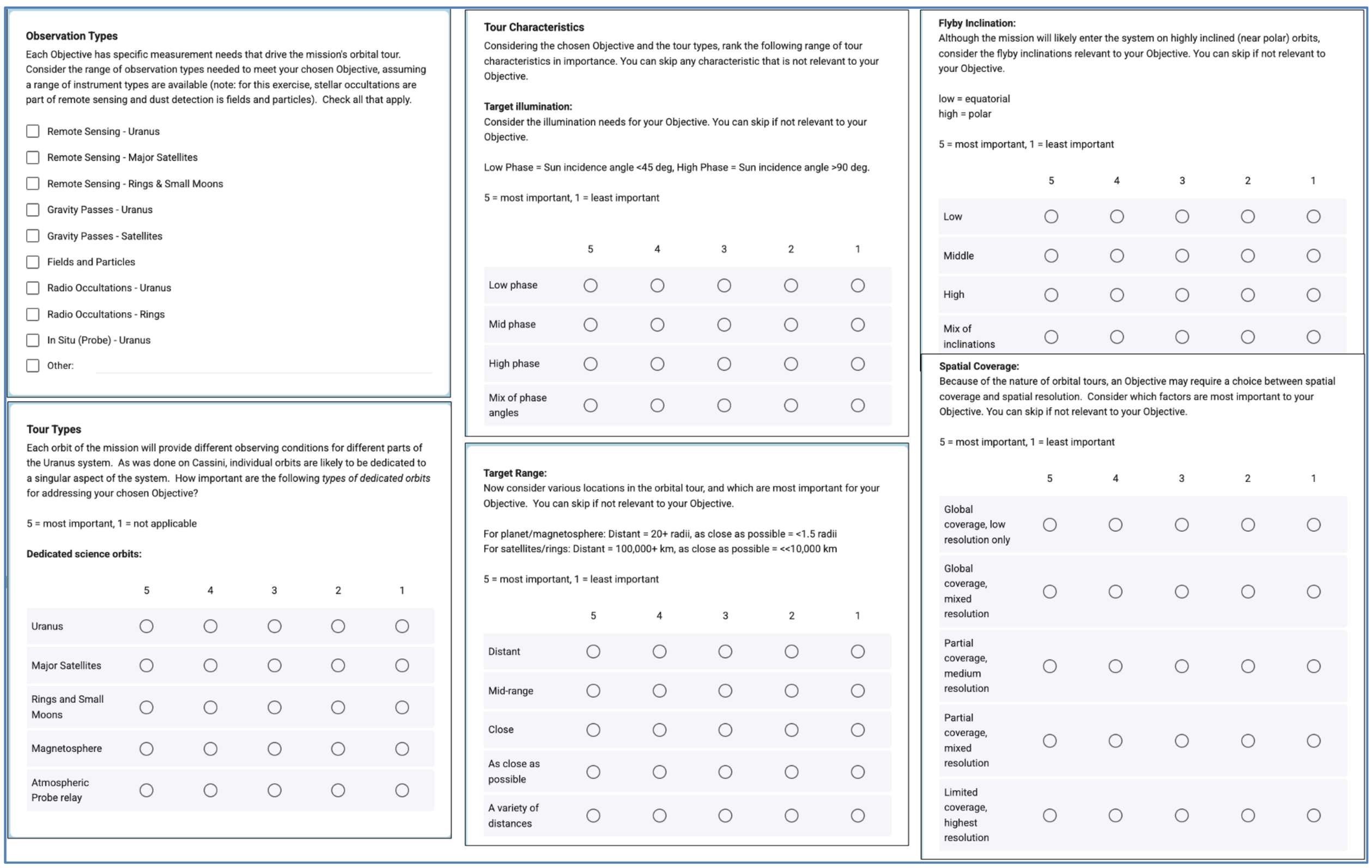

**Observation Types**

Each Objective has specific measurement needs that drive the mission's orbital tour. Consider the range of observation types needed to meet your chosen Objective, assuming a range of instrument types are available (note: for this exercise, stellar occultations are part of remote sensing and dust detection is fields and particles). Check all that apply.

- ☐ Remote Sensing - Uranus
- ☐ Remote Sensing - Major Satellites
- ☐ Remote Sensing - Rings & Small Moons
- ☐ Gravity Passes - Uranus
- ☐ Gravity Passes - Satellites
- ☐ Fields and Particles
- ☐ Radio Occultations - Uranus
- ☐ Radio Occultations - Rings
- ☐ In Situ (Probe) - Uranus
- ☐ Other:

**Tour Types**

Each orbit of the mission will provide different observing conditions for different parts of the Uranus system. As was done on Cassini, individual orbits are likely to be dedicated to a singular aspect of the system. How important are the following *types of dedicated orbits* for addressing your chosen Objective?

5 = most important, 1 = not applicable

**Dedicated science orbits:**

| | 5 | 4 | 3 | 2 | 1 |
|---|---|---|---|---|---|
| Uranus | ○ | ○ | ○ | ○ | ○ |
| Major Satellites | ○ | ○ | ○ | ○ | ○ |
| Rings and Small Moons | ○ | ○ | ○ | ○ | ○ |
| Magnetosphere | ○ | ○ | ○ | ○ | ○ |
| Atmospheric Probe relay | ○ | ○ | ○ | ○ | ○ |

**Tour Characteristics**

Considering the chosen Objective and the tour types, rank the following range of tour characteristics in importance. You can skip any characteristic that is not relevant to your Objective.

**Target illumination:**
Consider the illumination needs for your Objective. You can skip if not relevant to your Objective.

Low Phase = Sun incidence angle <45 deg, High Phase = Sun incidence angle >90 deg.

5 = most important, 1 = least important

| | 5 | 4 | 3 | 2 | 1 |
|---|---|---|---|---|---|
| Low phase | ○ | ○ | ○ | ○ | ○ |
| Mid phase | ○ | ○ | ○ | ○ | ○ |
| High phase | ○ | ○ | ○ | ○ | ○ |
| Mix of phase angles | ○ | ○ | ○ | ○ | ○ |

**Target Range:**
Now consider various locations in the orbital tour, and which are most important for your Objective. You can skip if not relevant to your Objective.

For planet/magnetosphere: Distant = 20+ radii, as close as possible = <1.5 radii
For satellites/rings: Distant = 100,000+ km, as close as possible = <<10,000 km

5 = most important, 1 = least important

| | 5 | 4 | 3 | 2 | 1 |
|---|---|---|---|---|---|
| Distant | ○ | ○ | ○ | ○ | ○ |
| Mid-range | ○ | ○ | ○ | ○ | ○ |
| Close | ○ | ○ | ○ | ○ | ○ |
| As close as possible | ○ | ○ | ○ | ○ | ○ |
| A variety of distances | ○ | ○ | ○ | ○ | ○ |

**Flyby Inclination:**
Although the mission will likely enter the system on highly inclined (near polar) orbits, consider the flyby inclinations relevant to your Objective. You can skip if not relevant to your Objective.

low = equatorial
high = polar

5 = most important, 1 = least important

| | 5 | 4 | 3 | 2 | 1 |
|---|---|---|---|---|---|
| Low | ○ | ○ | ○ | ○ | ○ |
| Middle | ○ | ○ | ○ | ○ | ○ |
| High | ○ | ○ | ○ | ○ | ○ |
| Mix of inclinations | ○ | ○ | ○ | ○ | ○ |

**Spatial Coverage:**
Because of the nature of orbital tours, an Objective may require a choice between spatial coverage and spatial resolution. Consider which factors are most important to your Objective. You can skip if not relevant to your Objective.

5 = most important, 1 = least important

| | 5 | 4 | 3 | 2 | 1 |
|---|---|---|---|---|---|
| Global coverage, low resolution only | ○ | ○ | ○ | ○ | ○ |
| Global coverage, mixed resolution | ○ | ○ | ○ | ○ | ○ |
| Partial coverage, medium resolution | ○ | ○ | ○ | ○ | ○ |
| Partial coverage, mixed resolution | ○ | ○ | ○ | ○ | ○ |
| Limited coverage, highest resolution | ○ | ○ | ○ | ○ | ○ |

**Figure 1**. Poll questions about observation and tour types, and general tour characteristics.

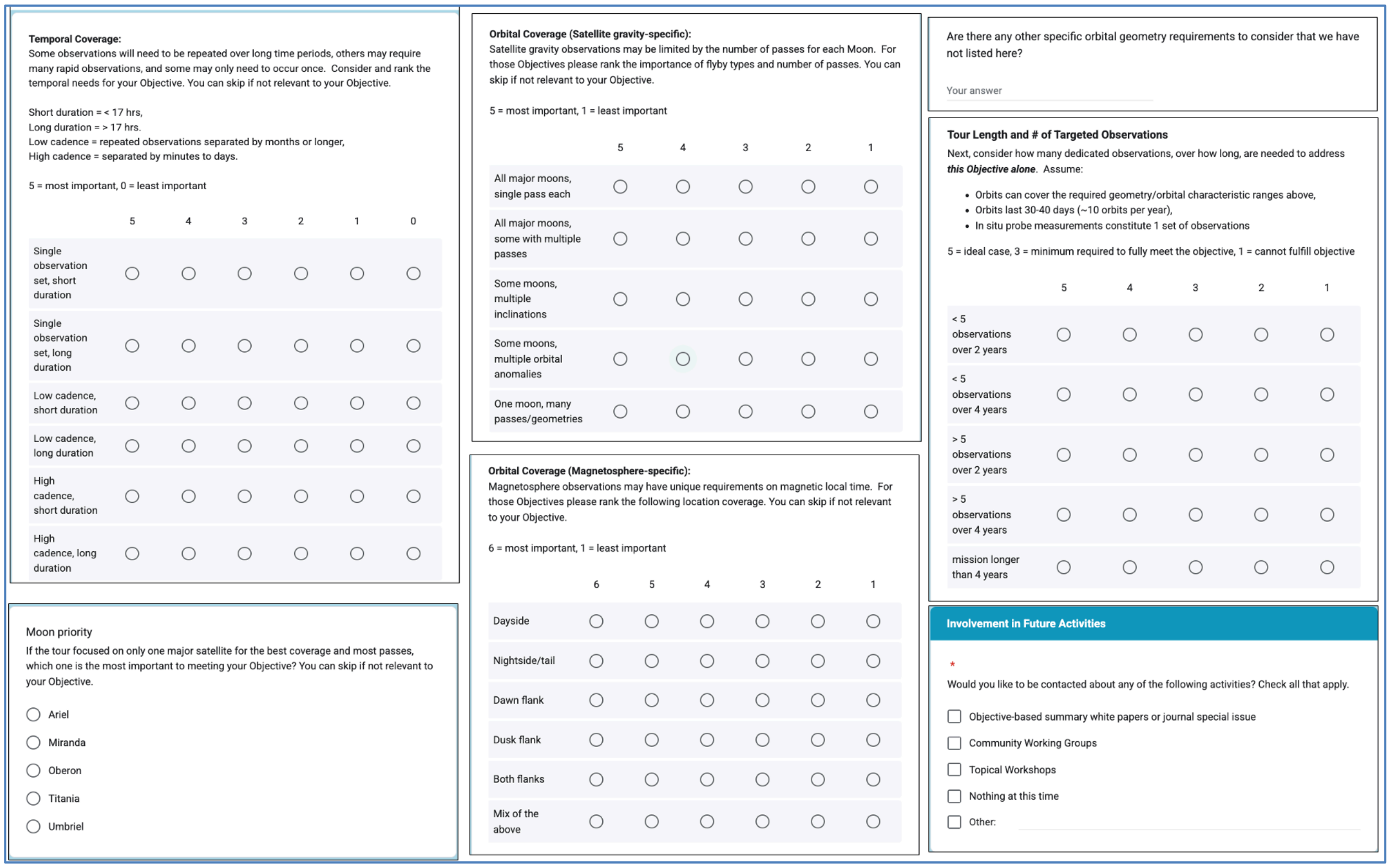

**Figure 2**. Poll questions containing specific tour geometry and coverage questions, as well as gauging interest in future activities.

**Aggregated Results by Objective**: All responses have been anonymized. For every Objective with more than six responses, results were aggregated individually. For Objectives with fewer than six responses, results were grouped with the closest scientific Objective to provide a more statistically significant tally. Thus, the following Objectives were combined: A3/I1 and I2 and I3; R1 and R2; M2 and M3; and S2 and S4. For each individual Objective/group, the numerical responses were tabulated and normalized by the number of responses to determine which tour characteristics were highest priority for that Objective. Figures 3 to 10 show the aggregated rankings of medium-, high-, and highest-priority primary tour characteristics for each Objective/group. It should be noted that not everyone responded to every question in their chosen Objective; in those cases, the votes for highest priority will not total to 100%. Areas where there were very few votes (i.e., low percentages in all options for a specific tour property) may perhaps indicate that there were no strong opinions on that particular item's priority for that Objective.

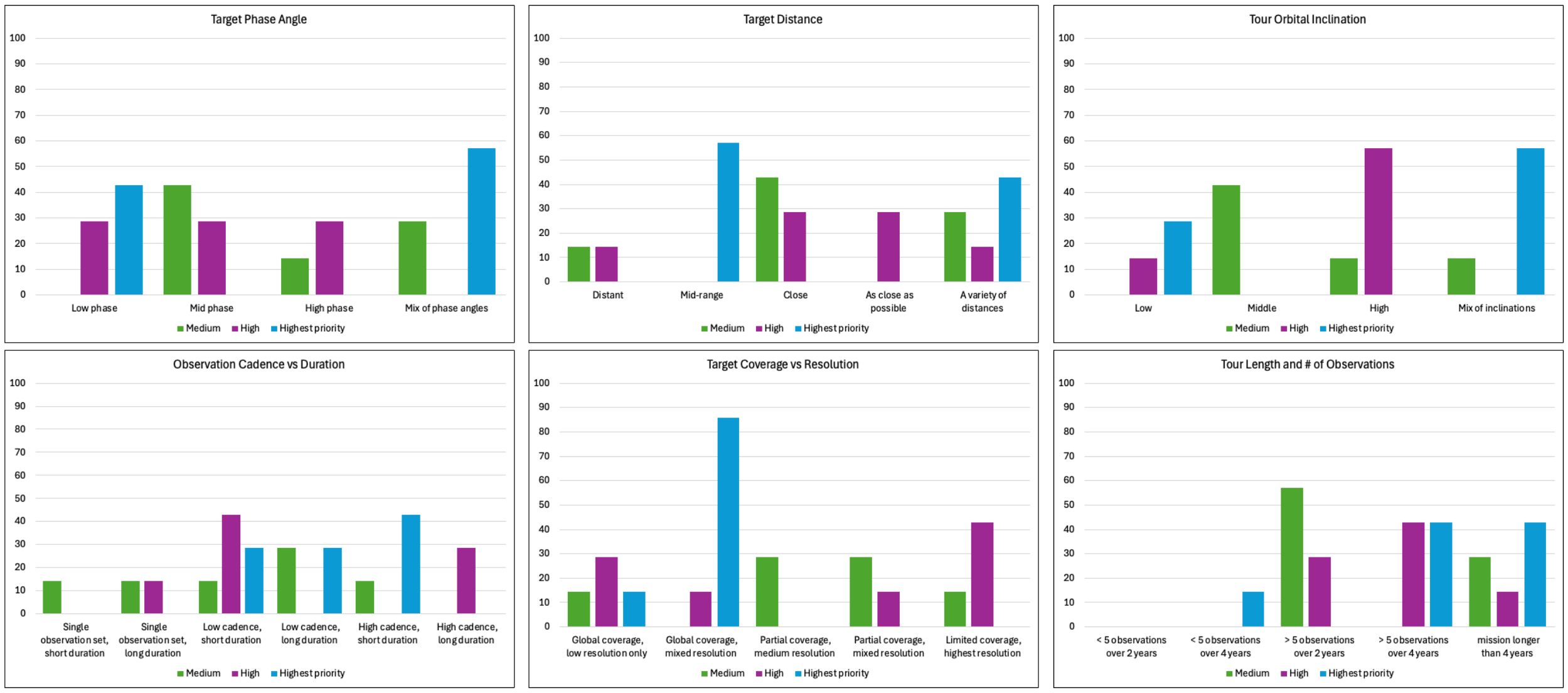

**Figure 3**. Objective A1 results.

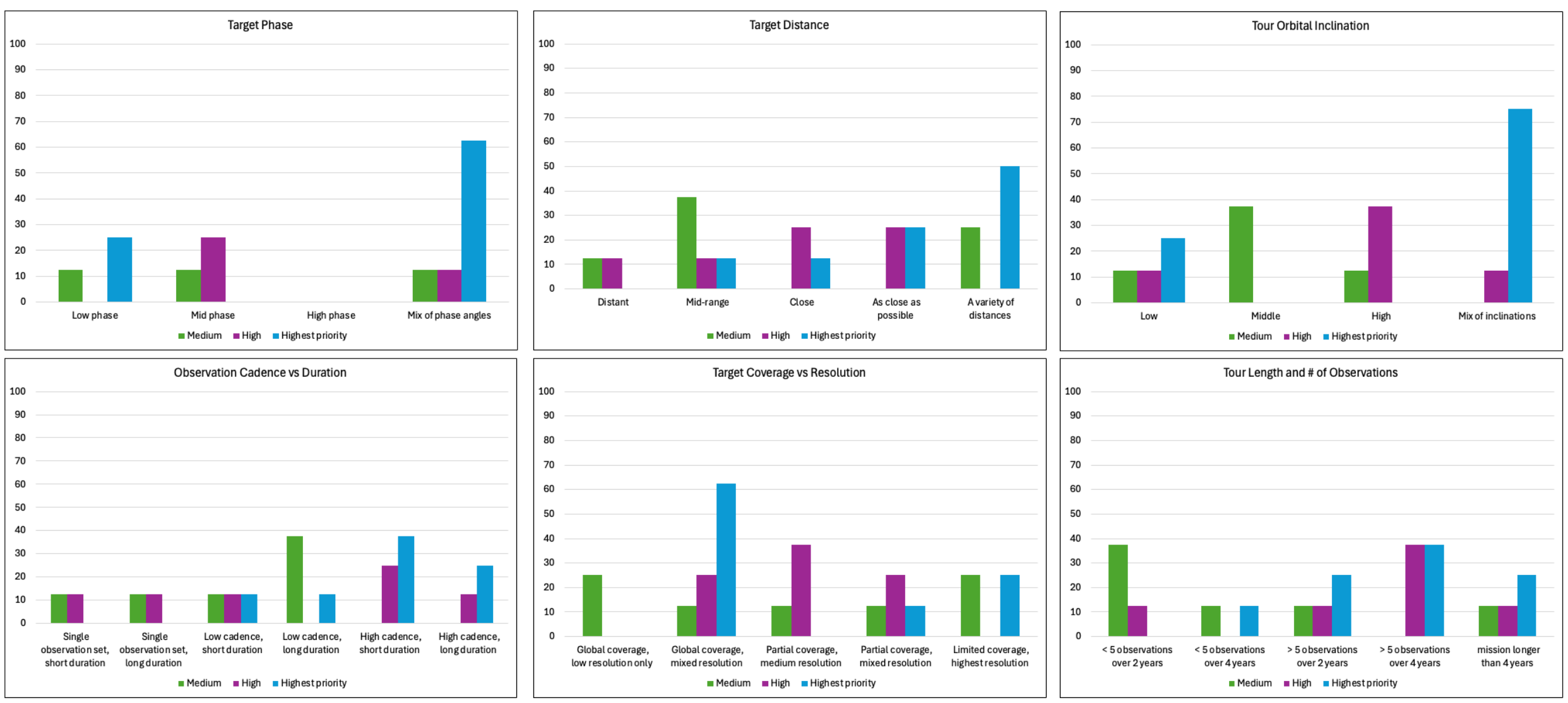

**Figure 4**. Objective A2 results.

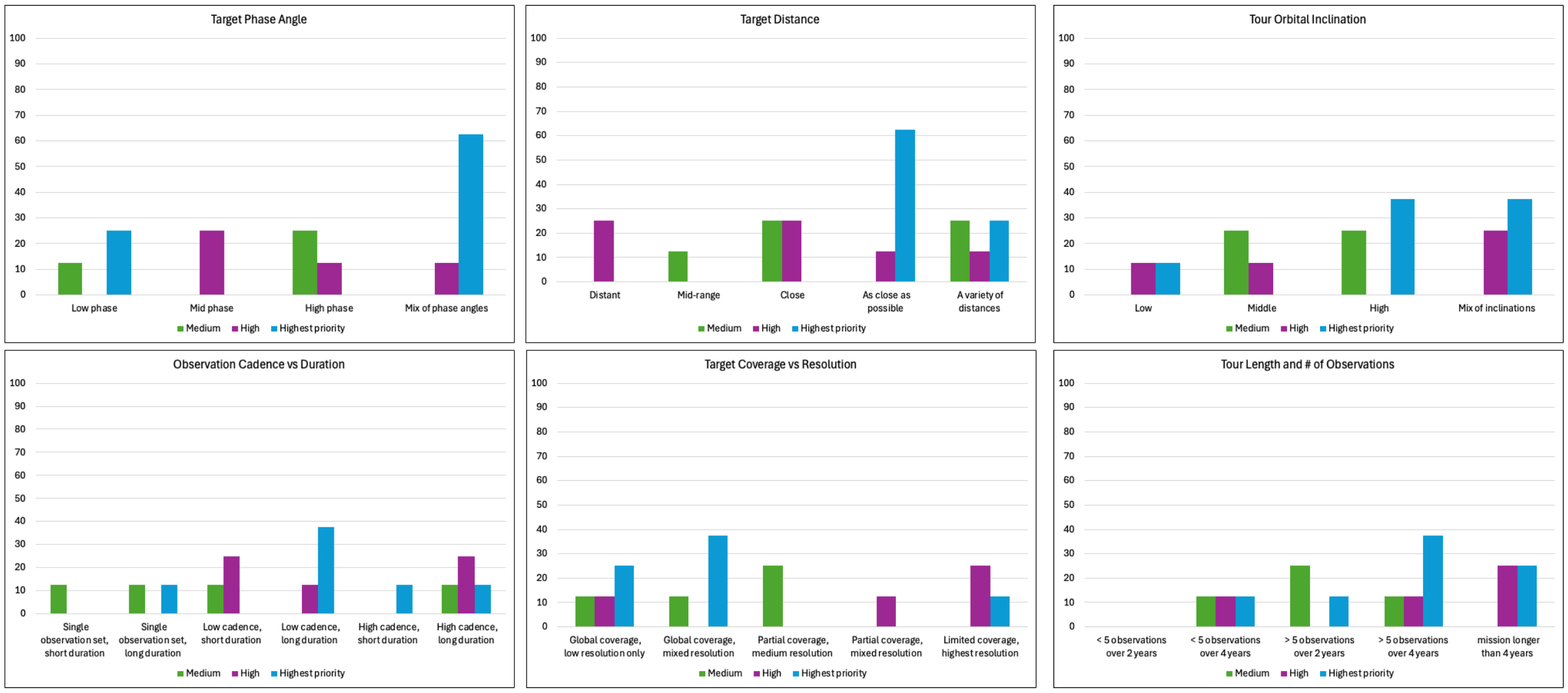

**Figure 5**. Objectives A3/I1, I2, I3 results.

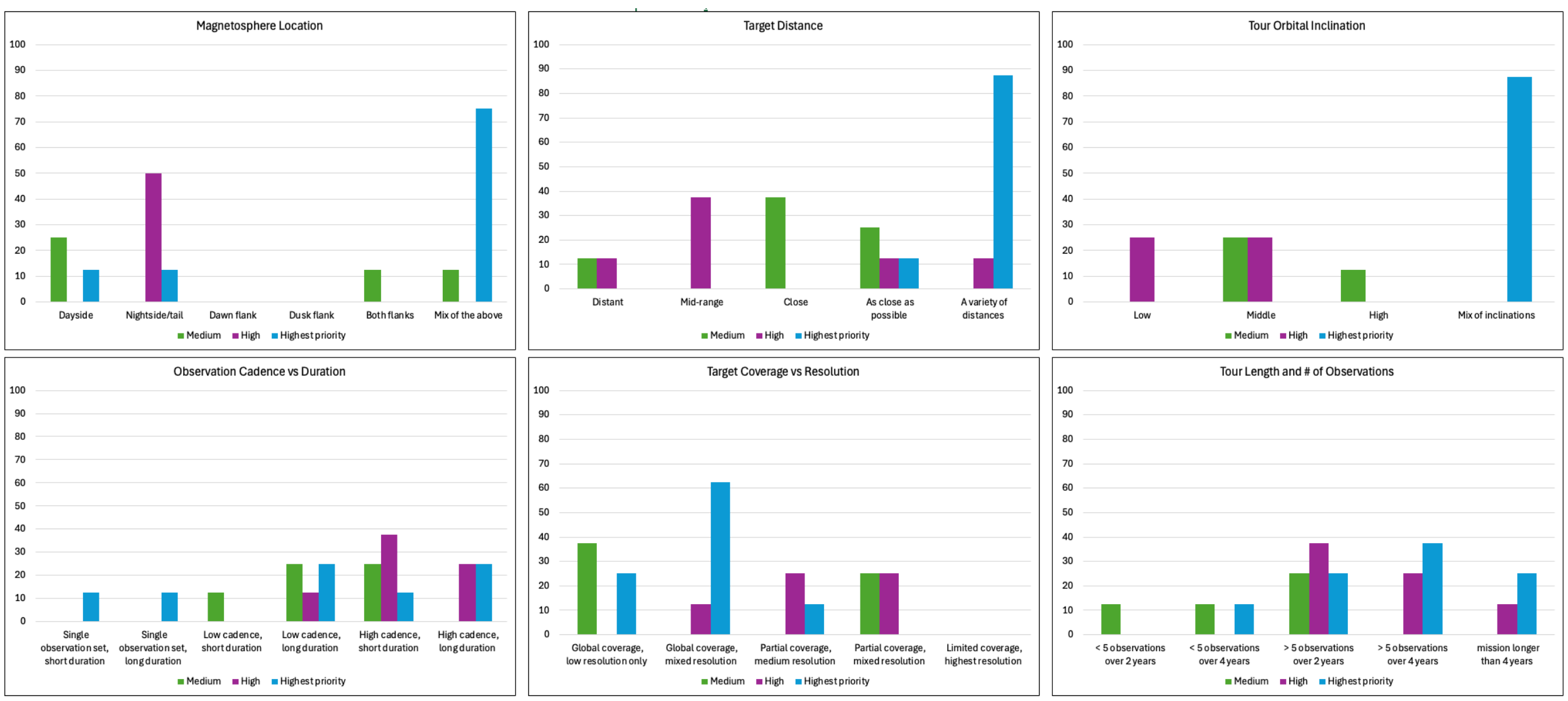

**Figure 6**. Objectives M2, M3 results.

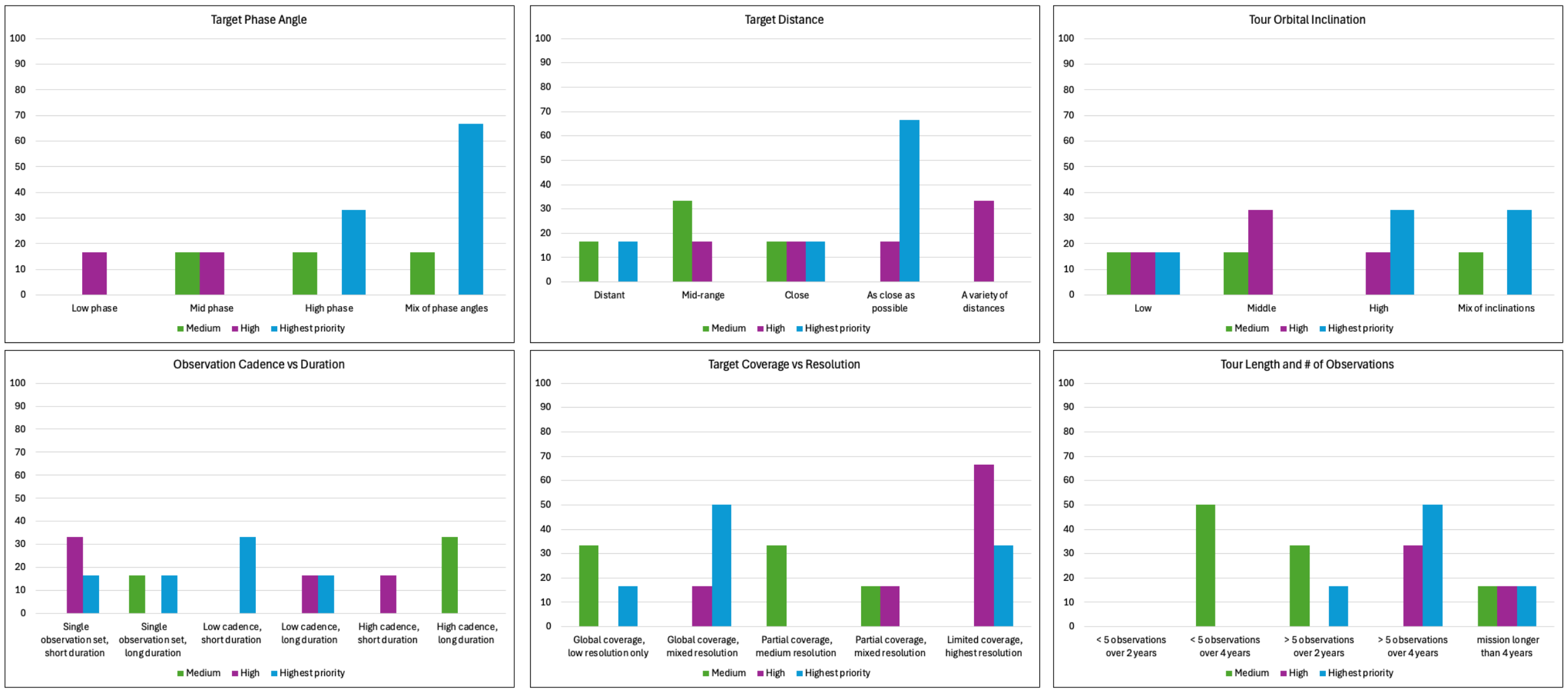

**Figure 7**. Objectives R1, R2 results.

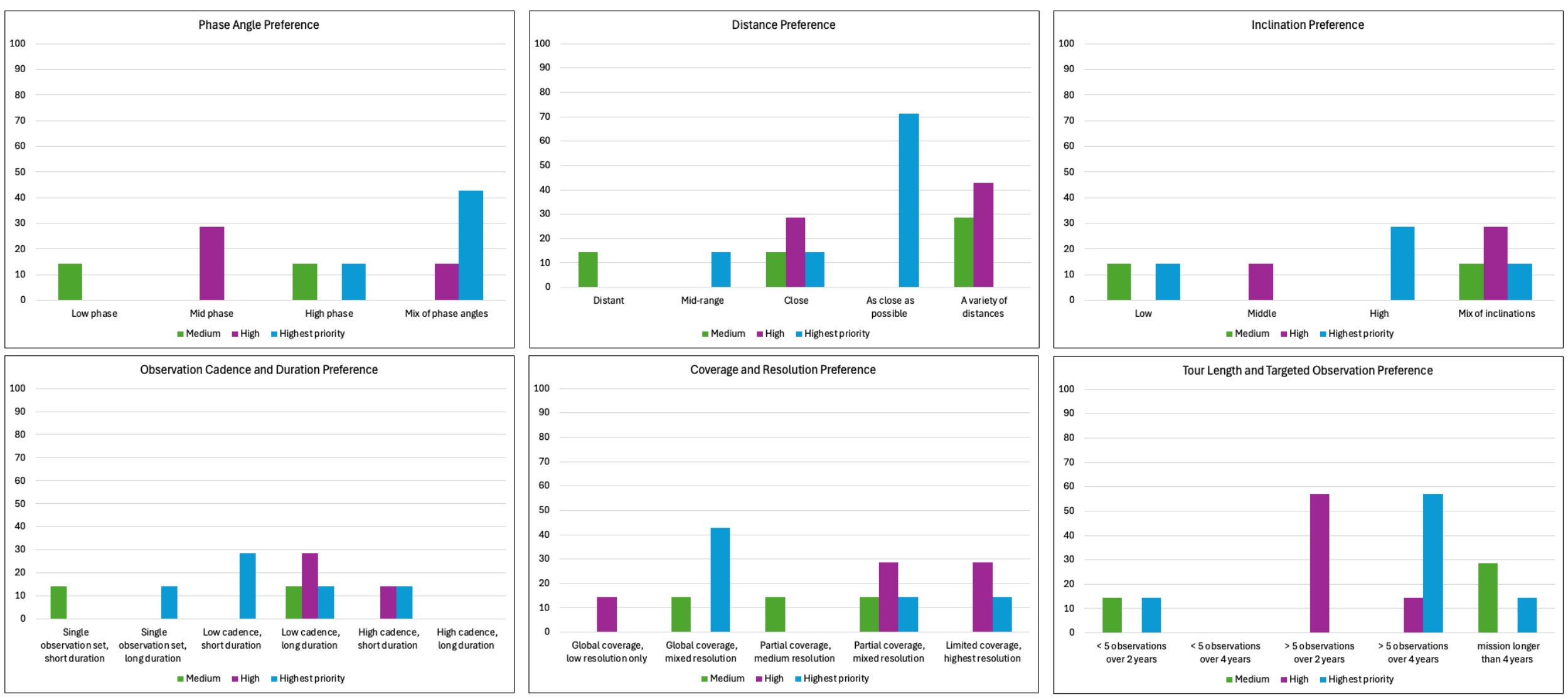

**Figure 8**. Objective S1 results.

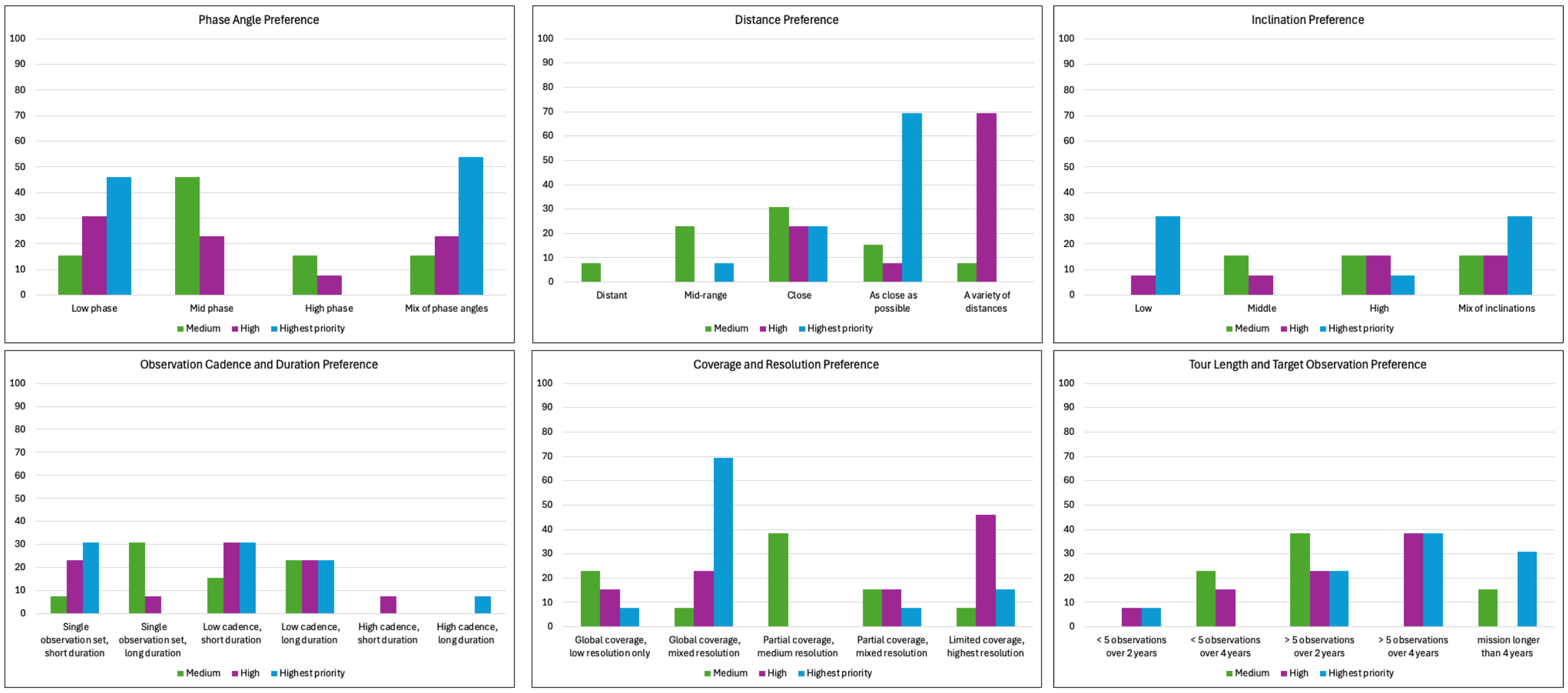

**Figure 9**. Objectives S2, S4 results.

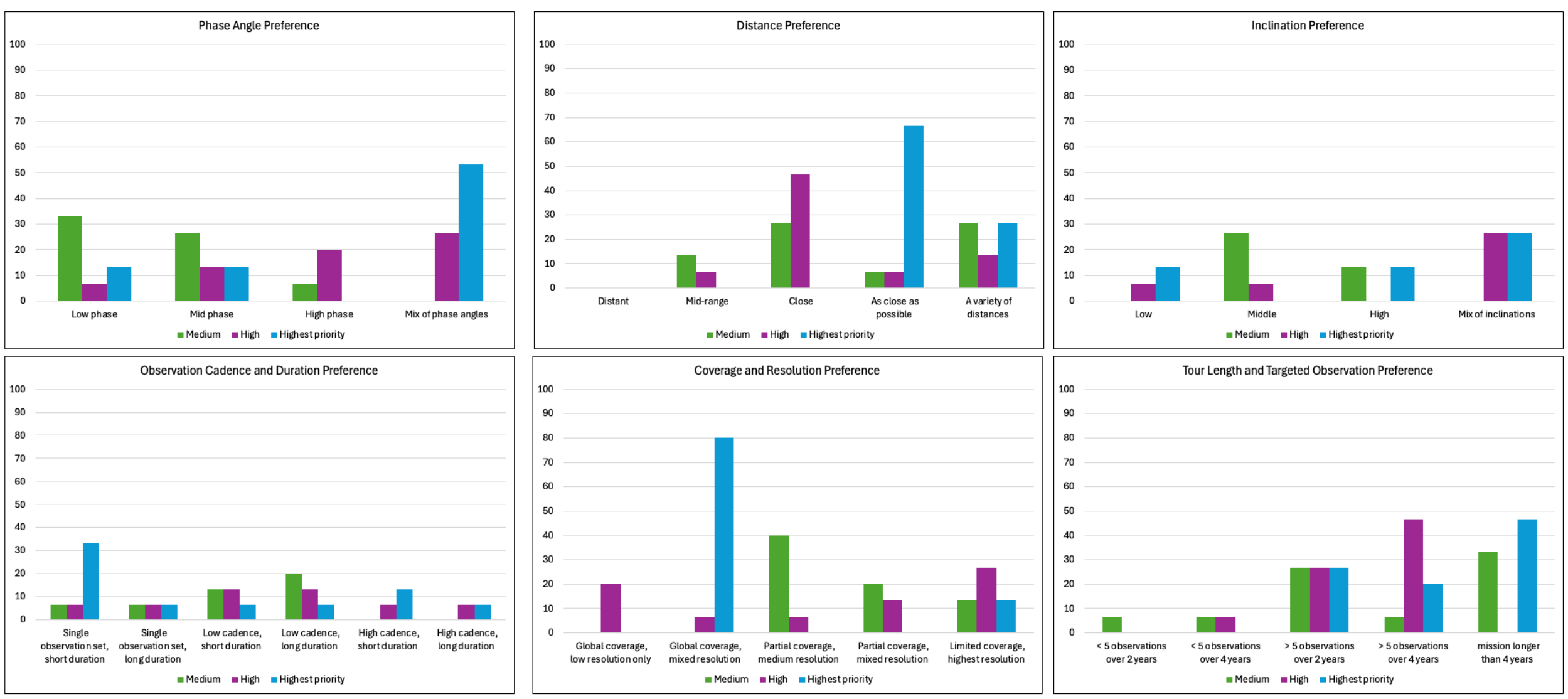

**Figure 10**. Objective S3 results.

**Cross-Objective Comparisons**: For the satellites it is informative to look across all the topical Objectives, in addition to those for individual Objectives, to determine what would make the best satellite tour. Figure 11 shows the tour coverage preference and the highest priority moon for each Objective. While respondents for all three Objectives show the preference for at least one flyby of each of the five major moons, Ariel is the highest priority target for detailed study, though Miranda is also of high interest.

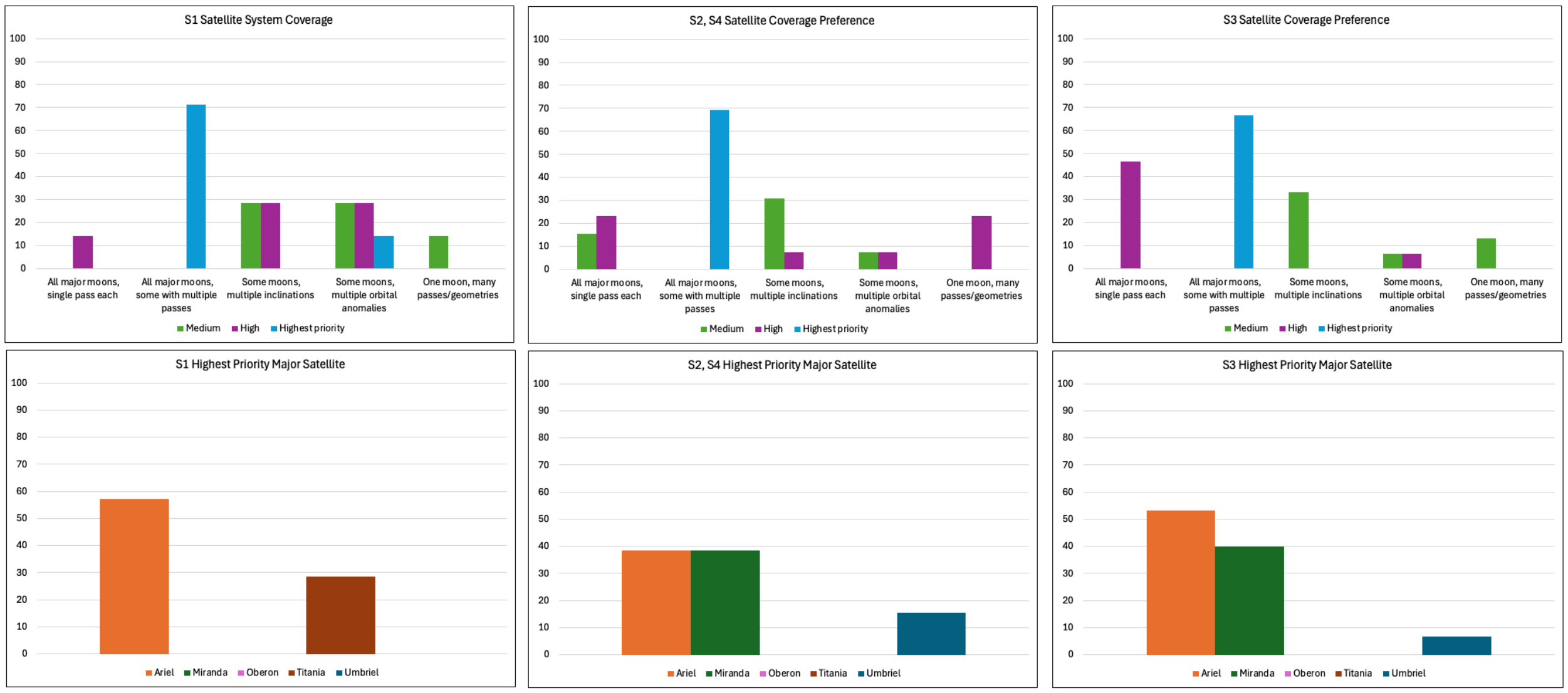

**Figure 11**. Comparative satellite results on moon coverage and priority.

Lastly, we compiled the highest priority tour characteristics across each Objective. This information will be useful for determining which tour might be best able to meet all Objectives simultaneously. A similar approach was used during the Cassini mission where different scientific disciplines were assigned orbit segments that best met their needs, and each discipline working group then interleaved instruments observations within allocated power and data volume constraints. In Figure 12, any tour characteristics identified by ≥ 50% of the respondents as their "highest" priority for that Objective are colored blue. If ≥ 50% of respondents ranked it as "high" or "highest" priority, it is pink. While there are no doubt nuances to this ranking, which will be of higher importance once the scientific instruments are selected, it is still informative to see if all disciplines' highest priorities can be achieved with a single tour.

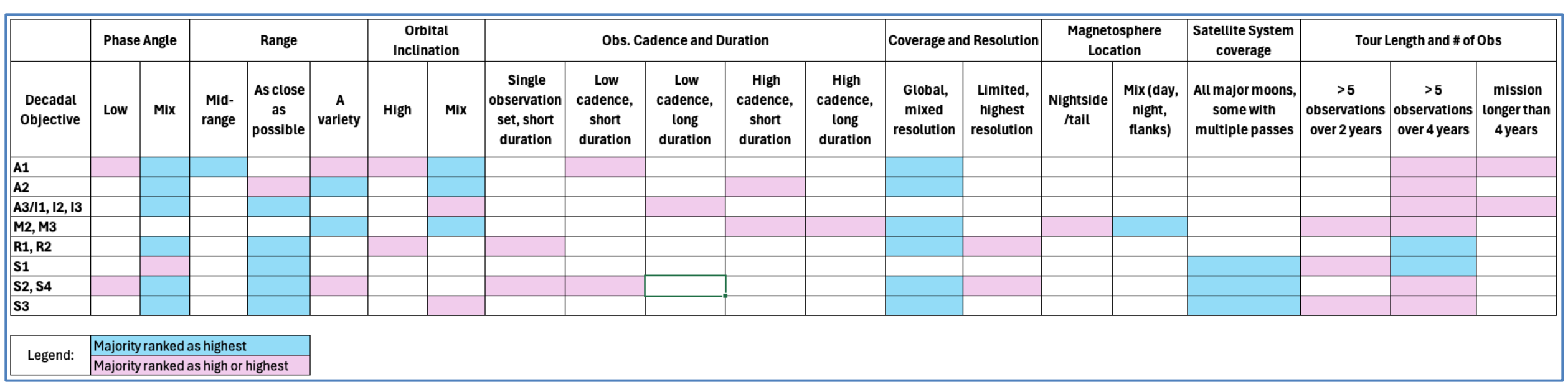

**Figure 12**. Highest priority tour characteristics by Objective.

**Data Availability**: The raw and aggregated can be downloaded from Zenodo (https://doi.org/10.5281/zenodo.15342504). The posted data have been anonymized, but some participants did supply their contact information as an indication of interest in participating in future activities. Please contact one of the authors if you would like the names of interested parties (by Objective) to participate in future white papers and/or other activities (see Figure 2 for categories).

**References:**

1. National Academies of Sciences, Engineering, and Medicine. 2023. *Origins, Worlds, and Life: Planetary Science and Astrobiology in the Next Decade*. Washington, DC: The National Academies Press. https://doi.org/10.17226/27209.
2. Simon & Nimmo et al. (2021). UOP Study Report: https://tinyurl.com/2p88fx4f
3. Landau et al. (2023). https://www.researchgate.net/publication/386106230_Trajectory_Options_for_a_Uranus_Orbiter_and_Probe
4. Ellison et al. (2024). https://icegiants.jhuapl.edu/uploadedDocs/events/presentations/Day2_Session4_5_Ellison_Trajectory.pdf